\documentclass[aps,twocolumn,showpacs]{revtex4-1}
\usepackage{float}
\usepackage{makeidx}
\usepackage{amsmath,amssymb,graphics,epsfig,color,times,bbm}

\begin{document}

\title{Quantum interferometer combining squeezing and parametric
amplification}
\author{Xiaojie Zuo$^{1}$}
\author{Zhihui Yan$^{1,2}$}
\email{zhyan@sxu.edu.cn}
\author{Yanni Feng$^{1}$}
\author{Jingxu Ma$^{1}$}
\author{Xiaojun Jia$^{1,2}$}
\email{jiaxj@sxu.edu.cn}
\author{Changde Xie$^{1,2}$}
\author{Kunchi Peng$^{1,2}$}

\affiliation{$^{1}$State Key Laboratory of Quantum Optics and Quantum Optics Devices,
Institute of Opto-Electronics, Shanxi University, Taiyuan, 030006, P. R.
China \\
$^{2}$Collaborative Innovation Center of Extreme Optics, Shanxi University,
Taiyuan 030006, P. R. China }

\begin{abstract}
High precision interferometers are the building blocks of precision
metrology and the ultimate interferometric sensitivity is limited by the
quantum noise. Here we propose and experimentally demonstrate a compact
quantum interferometer involving two optical parametric amplifiers and the
squeezed states generated within the interferometer are directly used for
the phase-sensing quantum state. By both squeezing shot noise and
amplifying phase-sensing intensity the sensitivity improvement of $4.86\pm
0.24$ dB beyond the standard quantum limit is deterministically realized and
a minimum detectable phase smaller than that of all present interferometers
under the same phase-sensing intensity is achieved. This interferometric
system has significantly potential applications in a variety of measurements
for tiny variances of physical quantities.
\end{abstract}

\maketitle

Metrology underpins the quantitative science and the improvement of
measurement precision leads to not only extensive detailed knowledge but
also new fundamental understanding of nature. The classical interferometer
consisting of linear beam splitters for optical beam splitting and
recombination is a powerful metrology tool and the phase change of light in
the interferometer is quite sensitive to a variety of variances of physical
quantities influencing the optical path, such as biological samples \cite%
{Bowen1}, continuous force and displacement \cite{Schliesser1}. Recently,
the gravitational-wave signals from the mergers of two binary black holes
and neutron stars have been observed by kilometer-scale laser
interferometers \cite{LIGO1}. However, the sensitivity for current
interferometer is limited by the vacuum fluctuations of electromagnetic
field inside the interferometer, which is generally called the shot noise
limit (SNL): $\Delta \phi _{SNL}=1/\sqrt{N}$ \cite{Caves1,Shapiro1}. The SNL
is the limitation of the precision for a classical optical device because
the existence of shot noise impedes its further improvement.

Quantum metrology employs quantum resources to improve the measurement
precision for breakthrough of the classical precision limit \cite{rmp1,Chen1}%
. In quest for high precision measurement, it has been demonstrated that the
sensitivity determined by classically behaving states can be surpassed if
exotic quantum states are applied \cite{Pooser3}. A photon number maximally
entangled state (NOON state) has been applied in interferometer measurements
in which the phase signals have been enhanced by N times and the
sensitivities have been increased beyond the classical limit \cite%
{Steinberg1,Maccone1,Takeuchi1}. In the Bose-Einstein condensates, the
classical limit has been beaten by using the entangled states to cancel
quantum noise via quantum destructive interference \cite{Oberthaler1,Klempt1}%
. The vacuum fluctuations have been significantly suppressed by making the
use of the squeezed states and the sensitivities beyond the SNL have been
achieved \cite{Xiao1,Grangier1}. The squeezed state injection into the
interferometric gravitational-wave detectors to further improving
sensitivities is progressing \cite{Lam1,Mavalvala1,Schnabel1,LIGO4}. In
quantum mechanics, the Heisenberg uncertainty gives the ultimate limit of
sensitivity, which is named as the Heisenberg limit (HL): $\Delta \phi
_{HL}=1/N$ \cite{Holland1} and lots of efforts have been made to pursue the
HL \cite{Pryde,Walmsley1,Guo1,Wineland,Sun,Guo2}. Especially, it is possible
to reach the HL by driving the interferometers with the squeezed states in
principle \cite{Smerzi1,Dowling1}.

On the other hand, the interferometers with novel structures provide an
alternative avenue to achieve high precision phase sensing. The parametric
processing has been wildly adopted in construction of interferometers for
realizing quantum metrology \cite{Yurke1,Plick1,Ou5}. The four wave mixing
(FWM) instead of linear beam splitters have been used for optical beam
splitting and recombination to form a SU(1,1) interferometer with an
improvement in sensitivity, where the signal related to the phase change is
enhanced while the noise level is kept close to the SNL \cite{Ou4,Chekhova}.
Recently, based on utilizing the truncated SU(1,1) interferometers several
groups have also demonstrated the enhancement of sensitivity by the
amplification of the signal and the reduction of the quantum noise \cite%
{Lett2,Lett3,Novikova} and implemented the quantum-enhanced measurement of
microscopic cantilever displacement \cite{Pooser1,Pooser2}. In the Ref \cite%
{Lett2}, the two-mode squeezed state generated by FWM, is used as probe of
interferometer. The second nonlinear interaction in the SU(1,1) is replaced
with two balanced homodyne detections (BHDs), which is used to the direct
measurement of the phase-sensing fields. The joint quadratures improve the
phase sensitivity in the truncated SU(1,1) interferometers by both
amplifying the phase-sensing intensity and squeezing the shot noise below
the SNL. In the truncated SU(1,1) interferometer the phase-sensing field is
directly injected into the detectors, thus its intensity has to be limited
below the saturation power of detectors. In the Ref \cite{Martini1}, an
optical parametric amplifier (OPA) is inserted into the single-photon-based
interferometer in the presence of losses and the achievable interferometric
sensitivity based on the heralded single-photon probe is improved, and still
scales as $\sqrt{\mathbf{N}}$. As well-known, the OPA consisting of an
optical cavity with a $\chi ^{(2)}$ nonlinear crystal is a stable solid
quantum device to reduce shot noise of optical fields \cite%
{Villar1,Treps1,Pfister1,Peng3,Peng1,Peng2,Furusawa1,Andersen1}, with which
the highest squeezing of 15 dB to date is achieved \cite{Vahlbruch}. Due to
both favorable features of noise squeezing and signal amplification the OPA
should be a good quantum optical resource to be applied for constructing a
quantum interferometer with the sensitivity beyond the SNL. So far, the
deterministically experimental realization of phase sensing with high
precision is still a significant challenge in quantum metrology.

In this letter we propose and demonstrate a feasible approach to construct a
quantum interferometer by combining squeezing and parametric amplification.
For interferometric metrology, the phase-sensing intensity is associated
with the interferometric sensitivity and the higher phase-sensing intensity
allows the better interferometric sensitivity. However, the ultimate
limitation of sensitivity is quantum noise of the phase-sensing light. Thus,
to implement a precise interferometric measurement the phase-sensing light
with higher intensity and as low as possible noise is wanted. For achieving
both squeezing of shot noise and amplifying of phase-sensing intensity
within a Mach-Zehender (MZ) interferometer, two OPAs are placed in two arms
of the interferometer, respectively. The
squeezed states generated within the interferometer are utilized as the phase-sensing quantum states. Due to effectively
exploiting shot noise squeezing and parameter amplifying
features of OPAs, the sensitivity of the interferometer
is deterministically improved, and the sub-SNL scaling sensitivity
is achieved. The experimental results show that the squeezed
noise floor of the output signal optical beam from the interferometer is $%
5.57\pm 0.19$ dB below the SNL when the phase-sensing intensity is amplified
from 5 $\mu $W to 75.3 $\mu $W. An enhancement of $4.86\pm 0.24$ dB in the
signal to noise ratio (SNR) in comparison with the classical device are
measured. When the phase-sensing intensity is 75.3 $\mu $W, the calculated
shot noise spectral density is 6.20$\times $10$^{-8}$ /$\sqrt{Hz}$. Our
measurement results have reached the Heisenberg-scale precision under low
phase-sensing intensity. Using the presented system, only by simply
manipulating the OPA gain the optimal phase sensitivity can be achieved. In
the presented OPA-based quantum interferometer the squeezed state of light generated
by OPA inside the interferometer is used as the phase-sensing probe and
directly interacted with the measured sample, so the transmission losses is
significantly reduced. In
our interferometer the destructive interference
output of phase-sensing fields is selected as the signal fields measured by
BHDs. In this case, the measured intensities are low enough, thus the
problem of power saturation for detectors is overcome. Without the power limitation to signal fields of
BHDs the presented system can be used not only for the measurement of
microscopic phase-sensing intensity, but also for that of higher
phase-sensing intensity.

\begin{figure}[tbp]
\begin{center}
\includegraphics[width=8.6cm]{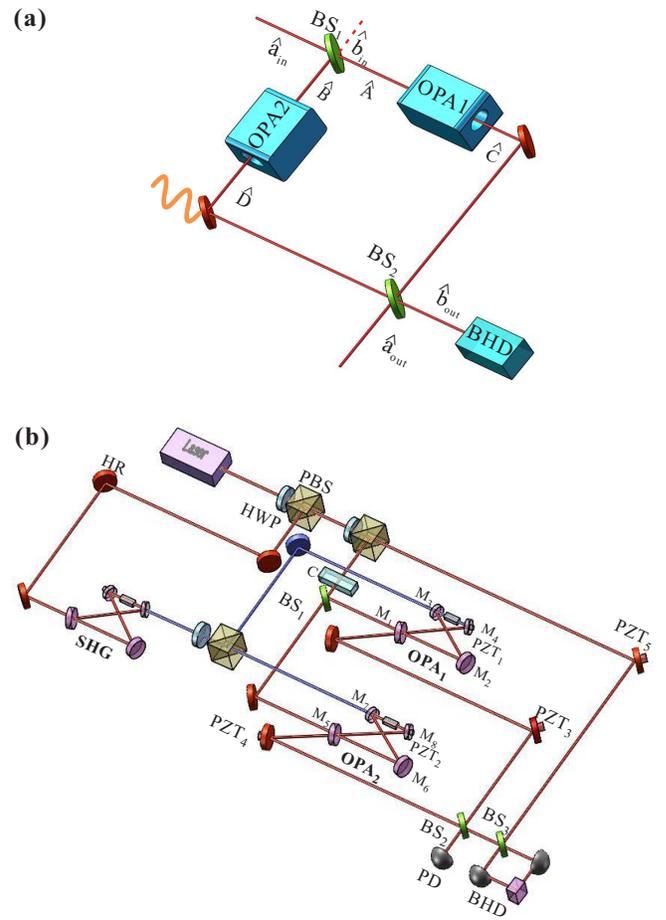}
\end{center}
\caption{(a) Schematic diagram for the quantum interferometer. The output
fields from OPAs are utilized as the phase-sensing light and the BHD
measures the quadrature phase of output optical field related to the phase
change $\protect\delta $. (b) Experimental setup for implementing the phase
change measurement using Mach-Zehender interferometer with two OPAs. The
phase change $\protect\delta $ is mimicked by the PZT$_{4}$. SHG: Second
harmonic generation. OPA$_{1,2}$: Optical parametric amplifier. BHD:
Balanced homodyne detection. BS$_{1-3}$: 50:50 beam splitter. HR:
high-reflection mirror. M$_{1-8}$: cavity mirror. HWP: Half-wave plate. PBS:
Polarization beam splitter. PZT$_{1-4}$: piezo-electric-transducer. C:
Chopper with attenuator.}
\end{figure}

Fig. 1 (a) is schematic diagram of a MZ interferometer involving two OPAs.
At first, the coherent laser $\hat{a}_{in}$ together with the vacuum state $%
\hat{b}_{in}$ are injected into the interferometer and splitted into two
modes $\hat{A}$ and $\hat{B}$ by the linear 50:50 beam splitter BS$_{1}$.
Then, $\hat{A}$ and $\hat{B}$ are amplified by two OPAs to be $\hat{C}$ and $%
\hat{D} $, respectively. The mode $\hat{D}$ passes through a sample to be
measured, which will give rise to a phase change $\delta $ of $\hat{D}$.
Next, the two beams $\hat{C}$ and $\hat{D}$\ are recombined by the linear
50:50 beam splitter BS$_{2}$ to produce the output modes $\hat{a}_{out}$ and
$\hat{b}_{out}$. The resultant interference signal $\hat{b}_{out}$ is
sensitive to the phase change $\delta $. Instead of measuring photon
intensity, we measure the quadrature phase $\hat{P}$ of $\hat{b}_{out}$ to
obtain the signal related to phase change $\delta $, which is implemented by
means of BHD. For a direct measurement processing the electronics noise will
swamp the weak signal, which is called as the dark count problem. In the
BHD, a strong local oscillator is employed to amplify the quadrature
components of weak sideband modes of a signal field. In this case the dark
count problem is overcome. While the relative phase between the optical
paths of two arms in the interferometer is kept to be $\pi +2k\pi $ ($k$ is
an integer) to obtain the
destructive interference, the quadrature phase $\hat{P}$ of output field $\hat{b}_{out}$
is detected by BHD, in which the phase change $\delta $ is recorded. The
sensitivity of the quantum interferometer is characterized by the
uncertainty of a single phase measurement, that is the minimum-detectable
phase shift $\Delta \phi $. The calculation details are given in the
supplemental material \cite{prl,Book,Tanimura,Jonathan1,Paris1,Goda} and we have the sensitivity of
interferometer in the lossless case as
\begin{equation}
\Delta \phi =\sqrt{\frac{\Delta ^{2}\hat{P}}{(\partial _{\phi }\hat{P})^{2}}}%
=\sqrt{\frac{(G-g)^{2}}{2(I_{ps}-g^{2})}},
\end{equation}%
where $\partial _{\phi }\hat{P}$\ is the change of quadrature phase $\hat{P}$
during the measurement with respect to a phase change $\delta $, $G\mathit{\
}$is the amplitude gain of OPA ($\left\vert G\right\vert ^{2}-\left\vert
g\right\vert ^{2}=1$) and $I_{ps}$ is the intensity of the phase-sensing
light ($I_{ps}=\frac{1}{2}(G+g)^{2}I_{0}+$\ $g^{2}$). The sensitivity can be
enhanced by a factor of 2$G$, when the phase-sensing intensity is larger
than the square of the gain factor $g^{2}$, which is large enough \cite{prl}. The
calculation details for the absolute value of the minimum detectable phase $%
\phi (\Omega )^{\min }$ of the quantum interferometer in the frequency
domain are shown in the supplemental material \cite{prl,Goda}

\begin{equation}
\phi (\Omega )^{\min }=\sqrt{\frac{4hce^{-2r}}{\lambda G^{^{\prime }}P_{in}}}%
,
\end{equation}%
where $h$ is the Plank constant, $c$ is the speed of light in vacuum, $%
\lambda $ is the laser wavelength, $P_{in}$ is the input optical intensity
of the quantum interferometer, $G^{^{\prime }}$ is the actual power gain
factor of the input light, and $r$ is the squeezing parameter associated
with shot noise reduction. It is noted that the minimal detectable phase $%
\phi (\Omega )^{\min }$ is independent of the analysis frequency.

The experimental setup for the OPA-based MZ interferometer is shown in Fig.
1 (b). A Ti:sapphire laser (Coherent MBR-110) with the output power of 2.5 W
pumped by a green laser (Yuguang DPSS FG-VIIIB) is used as the input signal
field of the quantum interferometer, fundamental field of second harmonic
generation (SHG) and local oscillation field of BHD. The input signal field
of the interferometer is splitted on the first linear 50:50 beam splitter BS$%
_{1}$ and the two optical beams from BS$_{1}$ are injected into two OPAs,
respectively. The output field from OPA$_{2}$ is modulated by the sinusoidal
signal of 2 MHz through the piezo-electric-transducer (PZT)$_{4}$ to mimic
the phase change and then is interfered with the output field from OPA$_{1}$
on the second linear 50:50 beam splitter BS$_{2}$. When the signal and noise
are measured, the stable bias phase of the interferometer is locked at $\pi
+2k\pi $\ ($k$\ is an integer) with the phase locking system based on the
Pound-Drever-Hall technique and a PZT$_{3}$\ mounted mirror. When the two OPAs are pumped by the vertically
polarized 448 nm continuous-wave single frequency laser from a SHG cavity  \cite{prl,Jia1},
the OPAs amplify the intensities of phase-sensing lights within the
interferometer and squeeze the noises on their phase-quadratures,
respectively. The BHD system consisting of a 50:50 beam
splitter BS$_{3}$, two photodiodes and a subtractor. Under the help of the
local oscillation light from the laser, the quadrature phase noise power of
the output field of interferometer is measured.

\begin{figure}[tbp]
\begin{center}
\includegraphics[width=8.6cm]{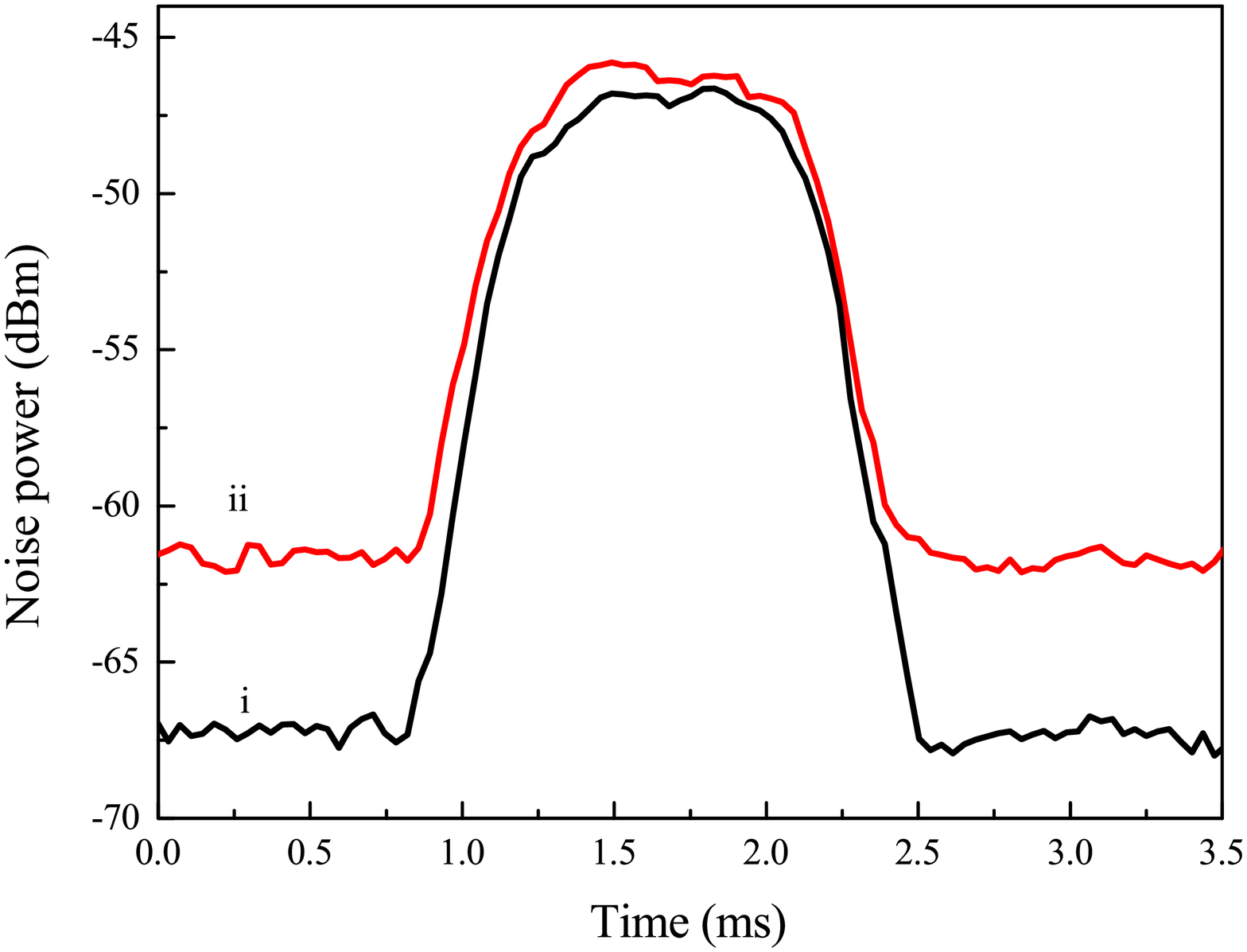}
\end{center}
\caption{The signal and noise levels measured at the output field of
OPA-based interferometer at analysis frequency of 2.0 MHz. The measurement
parameters are as follows for the spectrum analyzer : Resolution bandwidth:
100 kHz, Video bandwidth: 3 kHz, frequency span: 0.}
\end{figure}

The signal and noise levels of the output field of the OPA-based
interferometer at the analysis frequency of 2 MHz is shown in Fig. 2. When
the\ optical losses, the mode mismatch at BHD and other imperfection of the
interferometer are considered \cite{Hirano}, the expression of the
interferometric sensitivity should be dependent on the experimental
parameters, which is given by Eq. (12) in the supplemental material \cite{prl,Book,Tanimura}. The black trace (i) is the output noise power
measured at the case of two OPAs operating on the parametric amplification
with the OPA gain of 15. The SNL (the red trace (ii)) is measured when the
pump fields of two OPAs are blocked and a coherent state as the signal field
is injected. The reduction of the shot noise level below the SNL due to squeezing
makes the possibility of detecting tiny phase change submerged in the noise ocean. The squeezing of $5.57\pm 0.19$\ dB is the value measured on the output of
the OPA-based interferometer when the phase-sensing intensity is amplified
from 5 $\mu $W to 75.3 $\mu $W. When we implement the real measurement the signal light is slightly reduced by the unavoidable loss of 0.71 dB, thus the enhancement of SNR is naturally decreased to  $4.86\pm 0.24$\ dB in the comparison with the ideal SNL. Under the same phase-sensing intensity of
75.3 $\mu $W, the calculated shot noise spectral densities of quantum
interferometer and its corresponding SNL are 6.20$\times $10$^{-8}$ /$\sqrt{%
Hz}$ and 1.09$\times $10$^{-7}$ /$\sqrt{Hz}$, respectively (according to Eq.
(2)).

\begin{figure}[tbp]
\begin{center}
\includegraphics[width=8.6cm]{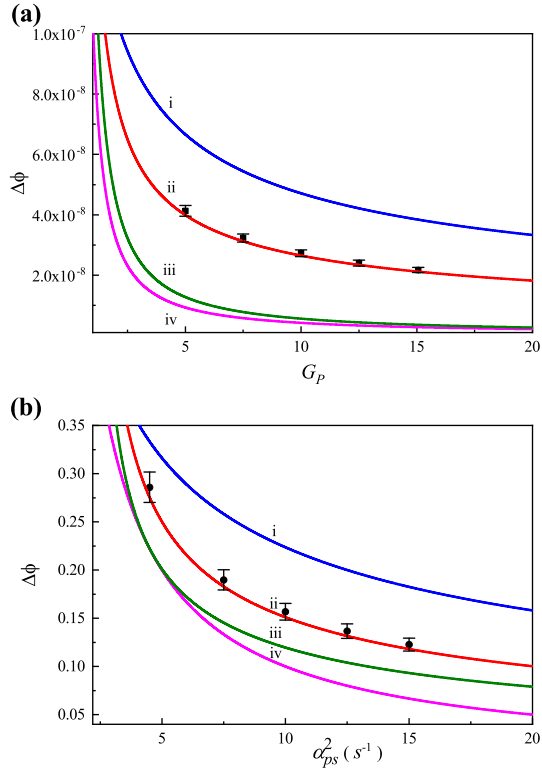}
\end{center}
\caption{(a) The effect of the OPA gain $G_{p}$ on the sensitivity of
quantum interferometer, when the input laser power of the interferometer is
10.0 $\protect\mu $W. (b) The phase sensitivity of quantum interferometer
versus the phase sensing intensity, in which the OPA gain $G_{p}$ of OPA is
5. The black squares and circles indicate the measured sensitivities.}
\end{figure}

The effect of the OPA gain $G_{p}$\ on the sensitivity of quantum
interferometer is shown in Fig. 3 (a), when the input laser power of the
interferometer is 10.0 $\mu $W (the corresponding $\alpha _{in}^{2}$ of
seeded light is 4.5$\times $10$^{13}$ s$^{-1}$). The blue trace (i)
corresponds to the SNL; the red trace (ii) and green trace (iii) define the
calculated sensitivities of the quantum interferometers when the parameters
are taken according to that at actually experimental (see Eq. (12) in the
supplementary material) and ideally lossless (see Eq. (1)) conditions,
respectively; the purple trace (iv) expresses the quantum Cram\'{e}r-Rao
bound (QCRB) of quantum interferometer (see Eq. (24) in the supplementary
material). It can be seen that the sensitivity is
improved with the increase of the OPA gain. The observed values of the black
squares are worse than the ideal values in trace (iii) because of the
influence of losses in the sub-quantum-limit interferometer. When the OPA
gain is 15, the corresponding value of $\Delta \phi _{SNL}$\ is 3.8$\times $%
10$^{-8}$ \cite{Ou5}. The calculated value of $\Delta \phi $\ can be
improved to 3.6$\times $10$^{-9}$\ in the lossless case, which is 10.6-fold
enhancement beyond the above SNL. In the lossless interferometer the
corresponding value of $\Delta \phi _{QCRB}$\ is 2.8$\times $10$^{-9}$\
obtained with the squeezing parameter r of 1.82 \cite{prl,Jonathan1,Paris1}. Therefore,
the interferometric phase sensitivity in the lossless case will be close to
the QCRB.

The sensitivities of quantum interferometer versus the $\alpha _{ps}^{2}$\
of the phase-sensing fields are illustrated in Fig. 3 (b), when the OPA gain
$G_{p}$\ is 5. The $\alpha _{ps}$\ in the horizontal axis stands for the
amplitude of the phase-sensing field, which relates to the phase-sensing
power $P_{ps}$\ by $\alpha _{ps}^{2}=\lambda P_{ps}/hc$. The blue Trace (i)
and purple trace (iv) are the SNL and the so-called HL, respectively. The
red trace (ii) is the calculated sensitivity of the quantum interferometers
in the experimental condition (see Eq. (12) in the supplementary material). The green trace (iii) defines the calculated sensitivity of the quantum
interferometers in the improved case when the losses are reduced to $%
L_{0}=0.002$, $\eta =0.99$ (see Eq. (12) in the supplementary material). In
the improved case with the phase-sensing intensity of 4.5 s$^{-1}$, the phase sensitivity is 0.22. The value of the so-called HL
calculated with the same intensity phase-sensing fields is also 0.22 \cite%
{Ou5}. Thus the quantum interferometer is possible to reach the sensitivity
allowed by the so-called HL.

In summary, we exploit two OPAs to construct a compact quantum
interferometer with a deterministically enhanced phase sensing. The tiny phase change submerged in the SNL can be measured due to both effects of amplified
phase-sensing intensity and squeezed noise. In the measurement with the low
phase-sensing intensity, the phase sensitivity has achieved the
Heisenberg-scale precision. The optical losses inside and outside
interferometer and the intracavity loss of OPA limit the measurement
precision of the present system. The reduction of these losses will enable
to obtain better phase-sensing ability. The quantum interferometer is
compatible with the SU(1,1) interferometer and squeezed state injection
systems, thus they can be combined together for the future sensitivity
improvement \cite{LIGO2,LIGO3}. Moreover, in the interferometric measurement
of the fragile samples, we have to utilize possibly low phase-sensing
intensity to protect samples from being damaged. In this case, the squeezed
states of light offer a liable option to directly measure tiny signals
submerged in the noise ocean. The wavelength used in our
interferometer is tunable around 895 nm, which matches not only cesium atom
\cite{Ma} but also biological tissue \cite{Zhu}. Our interferometer is
suitable for quantum biology sensing and spectroscopy. Besides the
application in the MZ interferometer, the method placing OPAs in
interferometer offers a potential to achieve the improvement of sensitivity
in other type interferometers with the quantum advantage of the OPA. The presented method promises the
unconditional quantum-enhanced precision metrology for any phase related
tiny signals.

This research was supported by the National Natural Science
Foundation of China (Grants No. 61925503, No. 61775127, No. 11654002, and
No. 11834010), the Key Project of the National Key R\&D
program of China (Grant No. 2016YFA0301402), the Program for Sanjin Scholars of Shanxi Province, and the
fund for Shanxi \textquotedblleft 1331 Project\textquotedblright\ Key
Subjects Construction.

\end{document}